\documentclass[aps,prl,twocolumn,superscriptaddress,amsfonts,amssymb,amsmath]{revtex4}
\usepackage{dcolumn}
\usepackage{fancyhdr}
\AtBeginDocument{

\def\equation{\begin{linenomath*}\begin{oldeqn}}
\def\endequation{\end{oldeqn}\end{linenomath*}}
}

\def\Public{true}
\usepackage{graphicx}% Include figure files 
\usepackage{dcolumn}% Align table columns on decimal point
\usepackage{amsmath}
\usepackage{bm}% bold math
\usepackage{slashed}
\usepackage{rotating}
\usepackage{subfigure}
\usepackage{psfrag}
\usepackage[multidot]{grffile}

\usepackage[tracking=true,kerning=true,expansion=true,spacing=false,factor=1100,stretch=20,shrink=20]{microtype}
\SetTracking{encoding={*}, shape=sc}{50} % add a little extra spacing between small-caps letters -- looks better

 \usepackage{siunitx}
\ifthenelse{\equal{\Public}{true}} {
\usepackage[unicode=true,pdfpagelabels=true,pdfusetitle,pdfauthor={The CDF Collaboration}]{hyperref}
\setkeys{Gin}{width=\linewidth,keepaspectratio}
}

\usepackage{xspace}

\newcommand{\PW}{\ensuremath{W}\xspace} 
\newcommand{\PZ}{\ensuremath{Z}\xspace}

\newcommand{\PWprime}{\ensuremath{W'}\xspace} 

\newcommand{\PnuR}{\ensuremath{\nu_R}\xspace}

\newcommand{\Pqt}{\ensuremath{t}\xspace}
\newcommand{\Pqb}{\ensuremath{b}\xspace}
\newcommand{\Pqc}{\ensuremath{c}\xspace}

\renewcommand{\to}{\ensuremath{\rightarrow}\xspace}
\newcommand{\ttbar}{\ensuremath{\Pqt\APqt}\xspace}

\newcommand{\ppbar}{\ensuremath{p\bar{p}}\xspace}

\newcommand{\wptb}{\ensuremath{\PWprime\to\Pqt\Pqb}\xspace}

\newcommand{\wplnu}{\ensuremath{\PWprime \to \ell \nu}\xspace}

\newcommand{\wptbxsbr}{\ensuremath{\sigma(\ppbar\to\PWprime)\times\mathcal{B}(\wptb)}\xspace}
\newcommand{\wptbbr}{\ensuremath{\mathcal{B}(\wptb)}\xspace}
\newcommand{\schannel}{\ensuremath{s}-channel\xspace}

\newcommand{\gev}{\ensuremath{\text{GeV}}\xspace}

\newcommand{\gevcc}{\ensuremath{\gev/c^2}\xspace}

\newcommand{\invfb}{\ensuremath{\text{fb}^{-1}}\xspace}

\newcommand{\met}{\ensuremath{\not\mathrel{\et}}\xspace}

\newcommand{\vecmet}{\ensuremath{\vec{\met}}\xspace}

\newcommand{\mpt}{\ensuremath{\not\mathrel{\pt}}\xspace}
\newcommand{\vecmpt}{\ensuremath{\vec{\mpt}}\xspace}

\newcommand{\mwp}{\ensuremath{M_{\PWprime}}\xspace}

\newcommand{\gwp}{\ensuremath{g_{\PWprime}}\xspace}
\newcommand{\gsm}{\ensuremath{g_{\text{SM}}}\xspace}

\newcommand{\nnqcd}{\ensuremath{\text{NN}_{\textsc{QCD}}}\xspace}

\newcommand{\btagonet}{\textsf{1T}\xspace}
\newcommand{\btagtl}{\textsf{TL}\xspace}
\newcommand{\btagtt}{\textsf{TT}\xspace}

\newcommand{\secvtx}{\texttt{SecVTX}\xspace}
\newcommand{\jetprob}{\texttt{JetProb}\xspace}

\newcommand{\atlas}{\textsf{ATLAS}\xspace}
\newcommand{\cms}{\textsf{CMS}\xspace}

\newcommand{\cdf}{\textsf{CDF}\xspace}
\newcommand{\cdftwo}{\textsf{CDF II}\xspace}

\newcommand{\dzero}{\textsf{D\O}\xspace}

\newcommand{\btagging}{\Pqb-tagging\xspace}
\newcommand{\btagged}{\Pqb-tagged\xspace}

\newcommand{\pythia}{\textsf{PYTHIA}\xspace}

\renewcommand{\secvtx}{\textsf{SecVTX}\xspace}
\renewcommand{\jetprob}{\textsf{JetProb}\xspace}

\renewcommand{\pythia}{\textsc{pythia}\xspace}

\renewcommand{\secvtx}{\textsc{secvtx}\xspace}
\renewcommand{\jetprob}{\textsc{jetprob}\xspace}

\renewcommand{\btagonet}{{1T}\xspace}
\renewcommand{\btagtl}{{TL}\xspace}
\renewcommand{\btagtt}{{TT}\xspace}
\renewcommand{\cdf}{{CDF}\xspace}
\renewcommand{\cdftwo}{{CDF II}\xspace}
\renewcommand{\dzero}{{D0}\xspace}
\renewcommand{\atlas}{{ATLAS}\xspace}
\renewcommand{\cms}{{CMS}\xspace}
\newcommand{\nnnonw}{NN$_{\mathrm{QCD}}$}
\newcommand{\nnnont}{NN$_{V\mathrm{jets}}$}
\newcommand{\nntt}{NN$_{t\bar{t}}$}
\newcommand{\nnsig}{NN$_{\mathrm{sig}}$}
\renewcommand{\ttbar}{\mbox{$t\bar{t}$}} % for some reason I get a "command already defined" for this

\newcommand{\Et}{\mbox{$E_T$}}
\newcommand{\Pt}{\mbox{$p_T$}}

\renewcommand{\met}{\mbox{$\protect \raisebox{0.3ex}{$\not$}\Et$}} % for some reason I get a "command already defined" for this
\renewcommand{\mpt}{\mbox{$\protect \raisebox{0.3ex}{$\not$}\Pt$}} % for some reason I get a "command already defined" for this

\renewcommand{\gevcc}{\mbox{GeV/$c^2$}} % for some reason I get a "command already defined" for this

\begin{document}

\ifthenelse{\equal{\Public}{true}} {
  \title
  {\texorpdfstring{\boldmath{Search for resonances decaying to top and bottom quarks with the CDF experiment
  %in $p\bar{p}$ collisions at $\sqrt{s}$ = 1.96 TeV
  }}{search for resonances decaying to a top and a bottom quark in $p\bar{p}$ collisions at $\sqrt{s}$ = 1.96 TeV}}
 }
 {
  \title{Search for New Resonances Decaying to a Top and a Bottom Quark in $p\bar{p}$ Collisions at $\sqrt{s}$ = 1.96 TeV}
 }

\affiliation{Institute of Physics, Academia Sinica, Taipei, Taiwan 11529, Republic of China}
\affiliation{Argonne National Laboratory, Argonne, Illinois 60439, USA}
\affiliation{University of Athens, 157 71 Athens, Greece}
\affiliation{Institut de Fisica d'Altes Energies, ICREA, Universitat Autonoma de Barcelona, E-08193, Bellaterra (Barcelona), Spain}
\affiliation{Baylor University, Waco, Texas 76798, USA}
\affiliation{Istituto Nazionale di Fisica Nucleare Bologna, \ensuremath{^{jj}}University of Bologna, I-40127 Bologna, Italy}
\affiliation{University of California, Davis, Davis, California 95616, USA}
\affiliation{University of California, Los Angeles, Los Angeles, California 90024, USA}
\affiliation{Instituto de Fisica de Cantabria, CSIC-University of Cantabria, 39005 Santander, Spain}
\affiliation{Carnegie Mellon University, Pittsburgh, Pennsylvania 15213, USA}
\affiliation{Enrico Fermi Institute, University of Chicago, Chicago, Illinois 60637, USA}
\affiliation{Comenius University, 842 48 Bratislava, Slovakia; Institute of Experimental Physics, 040 01 Kosice, Slovakia}
\affiliation{Joint Institute for Nuclear Research, RU-141980 Dubna, Russia}
\affiliation{Duke University, Durham, North Carolina 27708, USA}
\affiliation{Fermi National Accelerator Laboratory, Batavia, Illinois 60510, USA}
\affiliation{University of Florida, Gainesville, Florida 32611, USA}
\affiliation{Laboratori Nazionali di Frascati, Istituto Nazionale di Fisica Nucleare, I-00044 Frascati, Italy}
\affiliation{University of Geneva, CH-1211 Geneva 4, Switzerland}
\affiliation{Glasgow University, Glasgow G12 8QQ, United Kingdom}
\affiliation{Harvard University, Cambridge, Massachusetts 02138, USA}
\affiliation{Division of High Energy Physics, Department of Physics, University of Helsinki, FIN-00014, Helsinki, Finland; Helsinki Institute of Physics, FIN-00014, Helsinki, Finland}
\affiliation{University of Illinois, Urbana, Illinois 61801, USA}
\affiliation{The Johns Hopkins University, Baltimore, Maryland 21218, USA}
\affiliation{Institut f\"{u}r Experimentelle Kernphysik, Karlsruhe Institute of Technology, D-76131 Karlsruhe, Germany}
\affiliation{Center for High Energy Physics: Kyungpook National University, Daegu 702-701, Korea; Seoul National University, Seoul 151-742, Korea; Sungkyunkwan University, Suwon 440-746, Korea; Korea Institute of Science and Technology Information, Daejeon 305-806, Korea; Chonnam National University, Gwangju 500-757, Korea; Chonbuk National University, Jeonju 561-756, Korea; Ewha Womans University, Seoul, 120-750, Korea}
\affiliation{Ernest Orlando Lawrence Berkeley National Laboratory, Berkeley, California 94720, USA}
\affiliation{University of Liverpool, Liverpool L69 7ZE, United Kingdom}
\affiliation{University College London, London WC1E 6BT, United Kingdom}
\affiliation{Centro de Investigaciones Energeticas Medioambientales y Tecnologicas, E-28040 Madrid, Spain}
\affiliation{Massachusetts Institute of Technology, Cambridge, Massachusetts 02139, USA}
\affiliation{University of Michigan, Ann Arbor, Michigan 48109, USA}
\affiliation{Michigan State University, East Lansing, Michigan 48824, USA}
\affiliation{Institution for Theoretical and Experimental Physics, ITEP, Moscow 117259, Russia}
\affiliation{University of New Mexico, Albuquerque, New Mexico 87131, USA}
\affiliation{The Ohio State University, Columbus, Ohio 43210, USA}
\affiliation{Okayama University, Okayama 700-8530, Japan}
\affiliation{Osaka City University, Osaka 558-8585, Japan}
\affiliation{University of Oxford, Oxford OX1 3RH, United Kingdom}
\affiliation{Istituto Nazionale di Fisica Nucleare, Sezione di Padova, \ensuremath{^{kk}}University of Padova, I-35131 Padova, Italy}
\affiliation{University of Pennsylvania, Philadelphia, Pennsylvania 19104, USA}
\affiliation{Istituto Nazionale di Fisica Nucleare Pisa, \ensuremath{^{ll}}University of Pisa, \ensuremath{^{mm}}University of Siena, \ensuremath{^{nn}}Scuola Normale Superiore, I-56127 Pisa, Italy, \ensuremath{^{oo}}INFN Pavia, I-27100 Pavia, Italy, \ensuremath{^{pp}}University of Pavia, I-27100 Pavia, Italy}
\affiliation{University of Pittsburgh, Pittsburgh, Pennsylvania 15260, USA}
\affiliation{Purdue University, West Lafayette, Indiana 47907, USA}
\affiliation{University of Rochester, Rochester, New York 14627, USA}
\affiliation{The Rockefeller University, New York, New York 10065, USA}
\affiliation{Istituto Nazionale di Fisica Nucleare, Sezione di Roma 1, \ensuremath{^{qq}}Sapienza Universit\`{a} di Roma, I-00185 Roma, Italy}
\affiliation{Mitchell Institute for Fundamental Physics and Astronomy, Texas A\&M University, College Station, Texas 77843, USA}
\affiliation{Istituto Nazionale di Fisica Nucleare Trieste, \ensuremath{^{rr}}Gruppo Collegato di Udine, \ensuremath{^{ss}}University of Udine, I-33100 Udine, Italy, \ensuremath{^{tt}}University of Trieste, I-34127 Trieste, Italy}
\affiliation{University of Tsukuba, Tsukuba, Ibaraki 305, Japan}
\affiliation{Tufts University, Medford, Massachusetts 02155, USA}
\affiliation{University of Virginia, Charlottesville, Virginia 22906, USA}
\affiliation{Waseda University, Tokyo 169, Japan}
\affiliation{Wayne State University, Detroit, Michigan 48201, USA}
\affiliation{University of Wisconsin, Madison, Wisconsin 53706, USA}
\affiliation{Yale University, New Haven, Connecticut 06520, USA}

\author{T.~Aaltonen}
\affiliation{Division of High Energy Physics, Department of Physics, University of Helsinki, FIN-00014, Helsinki, Finland; Helsinki Institute of Physics, FIN-00014, Helsinki, Finland}
\author{S.~Amerio\ensuremath{^{kk}}}
\affiliation{Istituto Nazionale di Fisica Nucleare, Sezione di Padova, \ensuremath{^{kk}}University of Padova, I-35131 Padova, Italy}
\author{D.~Amidei}
\affiliation{University of Michigan, Ann Arbor, Michigan 48109, USA}
\author{A.~Anastassov\ensuremath{^{w}}}
\affiliation{Fermi National Accelerator Laboratory, Batavia, Illinois 60510, USA}
\author{A.~Annovi}
\affiliation{Laboratori Nazionali di Frascati, Istituto Nazionale di Fisica Nucleare, I-00044 Frascati, Italy}
\author{J.~Antos}
\affiliation{Comenius University, 842 48 Bratislava, Slovakia; Institute of Experimental Physics, 040 01 Kosice, Slovakia}
\author{F.~Anz\`a}
\affiliation{University of Oxford, Oxford OX1 3RH, United Kingdom}
\author{G.~Apollinari}
\affiliation{Fermi National Accelerator Laboratory, Batavia, Illinois 60510, USA}
\author{J.A.~Appel}
\affiliation{Fermi National Accelerator Laboratory, Batavia, Illinois 60510, USA}
\author{T.~Arisawa}
\affiliation{Waseda University, Tokyo 169, Japan}
% ...

\collaboration{CDF Collaboration}
\altaffiliation[With visitors from]{
\ensuremath{^{a}}University of British Columbia, Vancouver, BC V6T 1Z1, Canada,
\ensuremath{^{b}}Istituto Nazionale di Fisica Nucleare, Sezione di Cagliari, 09042 Monserrato (Cagliari), Italy,
\ensuremath{^{c}}University of California Irvine, Irvine, CA 92697, USA,
\ensuremath{^{d}}Institute of Physics, Academy of Sciences of the Czech Republic, 182~21, Czech Republic,
\ensuremath{^{e}}CERN, CH-1211 Geneva, Switzerland,
\ensuremath{^{f}}Cornell University, Ithaca, NY 14853, USA,
\ensuremath{^{g}}University of Cyprus, Nicosia CY-1678, Cyprus,
\ensuremath{^{h}}Office of Science, U.S. Department of Energy, Washington, DC 20585, USA,
\ensuremath{^{i}}University College Dublin, Dublin 4, Ireland,
\ensuremath{^{j}}ETH, 8092 Z\"{u}rich, Switzerland,
\ensuremath{^{k}}University of Fukui, Fukui City, Fukui Prefecture, Japan 910-0017,
\ensuremath{^{l}}Universidad Iberoamericana, Lomas de Santa Fe, M\'{e}xico, C.P. 01219, Distrito Federal,
\ensuremath{^{m}}University of Iowa, Iowa City, IA 52242, USA,
\ensuremath{^{n}}Kinki University, Higashi-Osaka City, Japan 577-8502,
\ensuremath{^{o}}Kansas State University, Manhattan, KS 66506, USA,
\ensuremath{^{p}}Brookhaven National Laboratory, Upton, NY 11973, USA,
\ensuremath{^{q}}Istituto Nazionale di Fisica Nucleare, Sezione di Lecce, Via Arnesano, I-73100 Lecce, Italy,
\ensuremath{^{r}}Queen Mary, University of London, London, E1 4NS, United Kingdom,
\ensuremath{^{s}}University of Melbourne, Victoria 3010, Australia,
\ensuremath{^{t}}Muons, Inc., Batavia, IL 60510, USA,
\ensuremath{^{u}}Nagasaki Institute of Applied Science, Nagasaki 851-0193, Japan,
\ensuremath{^{v}}National Research Nuclear University, Moscow 115409, Russia,
\ensuremath{^{w}}Northwestern University, Evanston, IL 60208, USA,
\ensuremath{^{x}}University of Notre Dame, Notre Dame, IN 46556, USA,
\ensuremath{^{y}}Universidad de Oviedo, E-33007 Oviedo, Spain,
\ensuremath{^{z}}CNRS-IN2P3, Paris, F-75205 France,
\ensuremath{^{aa}}Universidad Tecnica Federico Santa Maria, 110v Valparaiso, Chile,
\ensuremath{^{bb}}The University of Jordan, Amman 11942, Jordan,
\ensuremath{^{cc}}Universite catholique de Louvain, 1348 Louvain-La-Neuve, Belgium,
\ensuremath{^{dd}}University of Z\"{u}rich, 8006 Z\"{u}rich, Switzerland,
\ensuremath{^{ee}}Massachusetts General Hospital, Boston, MA 02114 USA,
\ensuremath{^{ff}}Harvard Medical School, Boston, MA 02114 USA,
\ensuremath{^{gg}}Hampton University, Hampton, VA 23668, USA,
\ensuremath{^{hh}}Los Alamos National Laboratory, Los Alamos, NM 87544, USA,
\ensuremath{^{ii}}Universit\`{a} degli Studi di Napoli Federico I, I-80138 Napoli, Italy
}
\noaffiliation
% Last update: $Date: 2015/01/15 01:38:35 $

\begin{abstract}
We report on a search for charged massive resonances decaying to top ($t$) and bottom ($b$) quarks in the full data set of proton-antiproton collisions at center-of-mass energy of $\sqrt{s} = 1.96$ TeV collected by the CDF~II detector at the Tevatron, corresponding to an integrated luminosity of 9.5 \invfb.
% We consider events having no identified charged lepton, a transverse energy imbalance, and two or three jets, of which at least one is consistent with originating from the decay of a \Pqb quark. 
% The reconstructed \PWprime transverse invariant mass distribution is used to discriminate between signal and SM background.
No significant excess above the standard model (SM) background prediction is observed. 
%Using a benchmark \wptb left-right symmetric model, 
We set 95\% Bayesian credibility mass-dependent upper limits on the heavy charged particle production cross section times branching ratio to $t b$. Using a SM extension with a \wptb and left-right-symmetric couplings as a benchmark model, we constrain the \wptb mass and couplings in the 300 to 900 GeV/$c^2$ range. 
%Assuming a \PWprime boson with SM-like couplings and allowed (forbidden) \PWprime decay to leptons, we exclude \wptb for \PWprime boson masses below 860 (880) \gevcsq. Relaxing the hypothesis on SM-like couplings, we exclude \PWprime boson coupling-strength values 
%$\gwp$ 
%as a function of \PWprime-boson mass above 10\% SM coupling strength for \mwp = 300 \gevcsq. 
The limits presented here are the most stringent for a charged resonance with mass  
in the range 300 -- 600 GeV/$c^2$ decaying to top and bottom quarks.

\end{abstract}


\maketitle 

% ======================================================================
%\section{Introduction}
% ======================================================================

Several modifications of the standard model (SM) of particle physics predict the existence of massive, short-lived states decaying to pairs of SM leptons or quarks.  Such a resonance decaying to a top ($t$) and a bottom ($b$) quark, $t  b$, appears in models such as left-right-symmetric SM extensions \cite{Pati:1974yy}, Kaluza-Klein extra dimensions \cite{Mimura:2002te, Burdman:2006gy}, technicolor \cite{Georgi:1989xz, Malkawi:1996fs} or little Higgs scenarios \cite{Perelstein:2005ka} featuring one or more massive charged vector bosons, generically denoted as \PWprime. Searches for \PWprime bosons in the \wptb decay channel are complementary to searches in the leptonic decay channel \wplnu, and probe the most general scenario where the couplings of the \PWprime boson to fermions are free parameters.

Recent searches in the \wptb channel have been performed by the \cdf \cite{Aaltonen:2009qu} and \dzero \cite{Abazov:2011xs} collaborations at proton-antiproton ($p \bar p$) collisions at 1.96\,TeV center-of-mass energy (CM) at the Tevatron, and by the \atlas \cite{Aad:2014xea} and \cms \cite{Chatrchyan:2014koa} collaborations in proton-proton collisions at 8\,TeV CM energy at the Large Hadron Collider (LHC). For mass scales approaching and surpassing 1\,TeV, the LHC experiments have superior sensitivity to the Tevatron experiments due to the enhancement of the production cross section at the higher center-of-mass energy of the collisions. 
%However, for the lower mass region the Tevatron experiments have competitive sensitivity for searches of particles produced in quark-initiated states where the dominant SM background processes are gluon-initiated\,\cite{Aaltonen:2012af, Aaltonen:2013kxa, CDF:2014uma}.
However, in the mass region well below 1\,TeV the Tevatron experiments have greater sensitivity due to the relative suppression of gluon-initiated backgrounds compared to the quark-initiated signals such as the one under consideration here.

In this Letter, we present a novel search for charged massive resonances decaying to $t b$ quark pair. The search is performed in events where the top quark decays to a $Wb$ pair and the \PW boson decays to a charged lepton and a neutrino; the two bottom quarks hadronize and produce two clusters of particles (jets).
% , but either the lepton is a electron or muon which is not identified, or is a hadronically decaying \Ptau reconstructed as a jet. 
% Events in this channel are characterized by the presence of significant transverse energy imbalance, two jets originated from \Pqb-quarks, and no identified charged leptons. 
Since no assumptions on the signal model other than on the natural width are made, this search is sensitive to any narrow resonant state decaying to a $t b$ final state.
A simple left-right symmetric SM extension \cite{PhysRevD.66.075011}, predicting the existence of \PWprime bosons of unknown mass and universal weak-coupling strength to SM fermions, is used as a benchmark model. The reconstructed width of the signal is dominated by resolution effects; the test signal is therefore applicable for any $W^{\prime}$-like particle whose width is small compared to the experimental resolution.

The data were collected at the Tevatron \ppbar collider at a center-of-mass energy of 1.96 TeV and were recorded by the \cdftwo detector~\cite{cdfiia}.
The detector consists of a silicon microstrip vertex detector and a cylindrical drift chamber immersed in a 1.4\,T magnetic field for vertex and charged-particle trajectory (track) reconstruction, surrounded by pointing-tower-geometry electromagnetic and hadronic calorimeters for energy measurement, and muon detectors outside the calorimeters\,\cite{cdfinfos}.
%\cdftwo uses a cylindrical coordinate system with azimuthal angle $\phi$, polar angle $\theta$ measured with respect to the positive $z$ direction along the proton beam, and the distance $r$ measured from the beamline. The pseudorapidity, transverse energy, and transverse momentum are defined as $\eta=-\ln\left[\tan(\frac{\theta}{2})\right]$, $E_{T}=E\sin{\theta}$, and $p_{T}=p\sin{\theta}$, respectively, where $E$ and $p$ are the energy and momentum of the outgoing particle. 
%The missing transverse energy \vecmet is defined by $\vecmet =-\sum_iE_T^i{\hat n}_i$, where ${\hat n}_i$ is a unit vector perpendicular to the beam axis that points to the $i^{th}$ calorimeter tower ($\met=|\vecmet|$). 

We analyze events accepted by the online event selection (trigger) that requires either the event missing transverse-energy $\slashed E_T$ to satisfy $\slashed E_T$$>$ 45 GeV or, alternatively $\slashed E_T$ $>$ 35 GeV and the presence of two or more jets, each with transverse energy $E_T$ $>$ 15 GeV. %The parametrization of the trigger efficiency~\cite{Potamianos:2011iwa} significantly improves the modeling of the trigger turn-on outside the fully efficient region, as verified using data control samples.  
The full data set 
%collected by the CDF experiment with the above triggers, and fully operational detector 
corresponds to an integrated luminosity of 9.5 fb$^{-1}$. Offline, we select events with $\slashed E_T$ $>$ 50 GeV, after correcting measured jet energies for instrumental effects~\cite{cdfjets}.  We further require events to have two or three high-$E_T$ jets, where the two jets $j_1, j_2$ with the largest transverse energies, $E_T^{j_1}$ and $E_T^{j_2}$, are required to satisfy $E_T^{j_1}$ $>$ 35 GeV and $E_T^{j_2}$ $>$ 25 GeV; the jet energies are determined from calorimeter deposits and corrected using charged-particle momentum measurements~\cite{Adloff:1997mi}. One leading jet is required to be within the silicon detector acceptance,  $|\eta|\ <\ 0.8$; the other satisfies $|\eta|\ <\ 2.0$.
%addition
In addition to the large missing transverse-energy indicating the presence of a high-$p_T$ neutrino, the presence of a $W$ boson decaying to an $e \nu_e$ or $\mu \nu_{\mu}$ pair is confirmed by requiring a reconstructed electron or muon. Leptonically decaying $\tau$ leptons are collected in the same way. Hadronically decaying $\tau$ leptons from the $W$ decay chain are mostly reconstructed as jets in the calorimeter. Three-jet events are thus retained, while events with more than three jets with $E_T^{j}$ $>$ 15 GeV and $|\eta|\ <\ 2.4$ are excluded.
%
%
%Taus from the $t\rightarrow Wb\rightarrow \tau\nu b$ decay chain appear either as soft electrons or muons c
%A fraction of the signal events is expected to produce tau ($\tau$) leptons from the $t\rightarrow Wb\rightarrow \tau\nu b$ decay. At CDF, hadronic $\tau$ leptons are mostly reconstructed as jets in the calorimeter; to open the acceptance to events with $\tau$ leptons, three jet events are included but events with more than three jets with $E_T^{j}$ $>$ 15 GeV and $|\eta|\ <\ 2.4$ are excluded.
\indent The majority of the background at this stage is quantum crhomodynamics (QCD) production of multijet events, which yields $\slashed E_T$ generated through jet-energy mismeasurements. Neutrinos produced in semileptonic $b$-hadron decays also contribute to the $\slashed E_T$ of these events. In both cases, the $\slashed{\vec{E}}_T$ is typically aligned with the projection on the transverse plane of the second or third jet momentum.
%$\vec{E}_\mathrm{T}^{j_2}$ (or $\vec{E}_\mathrm{T}^{j_{3}}$, for events with a third jet), 
Events are rejected by requiring the azimuthal separation $\Delta \varphi$ between $\slashed{\vec{E}}_T$ and $\vec{E}_\mathrm{T}^{j_{2}}$ (or $\vec{E}_\mathrm{T}^{j_{3}}$) to be larger than 0.4. 
%Events with four or more reconstructed jets, where each jet has transverse energy in excess of 15 GeV and pseudorapidity $|\eta|\ <\ 2.4$, are rejected.
The resulting sample, {\it pretag}, contains 391\,229 events, about 940 of these would originate from the decay of a 300 GeV/\,$c^2$ $W^{\prime}$ boson with SM-like couplings.

In order to identify jets originated from the hadronization of a \Pqb quark (``\Pqb-tagged''), we use two different algorithms, each tuned either for making a very pure selection (Tight), or for making a somewhat less pure selection that is more efficient (Loose). The \secvtx algorithm \cite{secvtx} looks for a vertex displaced from the collisions point produced by the in-flight decay of a $b$-flavored hadron; for this analysis we choose the tight (T) working point. 
%corresponding to an efficiency of approximately 40\,\%.  
The \jetprob algorithm \cite{jetprob} determines the probability that the tracks within a jet originate from the primary vertex; we choose for the latter algorithm the loose (L) working point.  The efficiency for each $b$-tagging algorithm is approximately 40--50\%. 
%The choice to use \secvtx and \jetprob insted of the \hobit tagger useÄad in Ref. \cite{schannelmetbb} is due to the fact that we expect in this search a large contribution from events with high energetic jets while \hobit is not able to distinguish light-flavor jets for heavy-flavor jets if the jet energy is larger than 200 GeV. 
We require at least one of the first two leading jets in $E_T$ to be tagged by the \secvtx algorithm. Events are further divided among twelve statistically independent subsamples, depending on whether there are no additional $b-$tagged jets (\btagonet), or an another jet is tagged by \jetprob but not by \secvtx (\btagtl), or tagged by \secvtx (\btagtt), the number of jets (two-jet or three-jet sample) and the presence or absence of a reconstructed electron or muon. 
%The latter subsample is sensitive to both un-identified electrons and muons, and to tau leptons. 
%The minimum $p_T$ requirement for the definition of the lepton is 20\,GeV.
This division increases sensitivity because signal-to-noise ratio and background composition differ across subsamples.
%At this stage of the analysis 483 signal events for a $W'$ boson mass of 300\,GeV/$c^2$, would be accepted by the selection, compared to a total number of data events of 25\,256.
The resulting {\it preselection} sample contains 25\,256 events, to which a $W'$ boson with SM-like couplings and 300 GeV/\,$c^2$ mass would contribute about 480 events.

% One of the leading sources of significant \met in production of multijet events due to QCD arises from the mismeasurement of jet energies. Additionally, neutrinos from semileptonic \Pqb decays can also produce significant \met in these events. In both cases, the \vecmet is often aligned with $\vec{E}_\mathrm{T}^{j_2}$, and such events are rejected by requiring \mbox{$\Delta\varphi(\vecmet,\vec{E}_\mathrm{T}^{j_1}) \geq 1.5$} and \mbox{$\Delta\varphi(\vecmet,\vec{E}_\mathrm{T}^{j_{2,3}}) \geq 0.4$}.

% We reject events with four reconstructed jets, where each jet exceeds the minimum transverse energy threshold ($E_T >15$ \gev) and has pseudorapidity $\abseta < 2.4$. 

% ======================================================================
%\section{Signal and Background Model}
% ======================================================================
The dominant contribution to the preselection sample is due to QCD multijet production. Other processes giving significant contributions are top-antitop quark-pair production (\ttbar), electroweak single-top-quark production, dibosons (${\it WW}$, ${\it WZ}$), and production of jets in association with a boson ({\it V}+jets, where {\it V} stands for a {\it W} or a {\it Z} boson), including both heavy-flavor jets (from \Pqb or \Pqc quarks) and jets from light-flavor quarks or gluons that have been erroneously \btagged. 

A combination of data and simulations making use of Monte Carlo (MC) integration are used to derive the estimates for SM background contributions.
 % The \alpgen generator \cite{alpgen} is used to simulate events for diboson and \wzjets; \ttbar and single top are simulated with \powheg \cite{powheg}. Parton showering is performed by \pythia \cite{pythia}. 
% The event generation process includes a simulation of the detector response \cite{geant}, and the resulting simulated event samples are subjected to the same reconstruction and analysis chain as the data. 
% The normalization for \ttbar, single top, and diboson templates is constrained to the theoretical cross section value. The normalization for \wzjets templates is left unconstrained in the final fit procedure.as
The kinematic distributions of events associated with top-quark pair, single top quark, {\it V}+jets, $W+c$, diboson ($VV$) and associated Higgs and {\it W} or {\it Z} boson ($VH$) production are modeled using simulated samples.
The {\sc alpgen} generator~\cite{alpgen} is used to model {\it V}+jets at leading order (LO) in the strong-interaction coupling with up to four partons produced at tree level, based on generator-to-reconstructed-jet matching~\cite{Mangano:2006rw,Alwall:2007fs}. 
The {\sc powheg}~\cite{powheg} generator is used to model $t$- and $s$-channel single top quark production, while {\sc pythia}~\cite{pythia} is used to model top-quark-pair, $VV$, and $VH$ production. Each event generator uses  the CTEQ5L parton distribution functions (PDF)~\cite{Lai:1999wy}. 
Parton showering is simulated in using {\sc pythia}. 
%, and tuned to the Tevatron underlying-event data~\cite{Aaltonen:2010rm}. 
Event modeling also includes simulation of the detector response using {\sc geant}~\cite{geant}. The simulated events are reconstructed and analyzed in the same way as the experimental data. Normalizations of the contributions from $t$- and $s$-channel single top quark, $VV$, $VH$, and $t\bar{t}$ pair production are taken from theoretical cross-sections~\cite{Kidonakis:2010tc,Campbell:1999ah,Baglio:2010um,Baernreuther:2012ws}, while the normalization for $W+c$ production is taken from the measured cross section~\cite{Aaltonen:2012wn}. 
For {\it V}+jets production, the heavy-flavor contribution is normalized based on the number of $b$-tagged events observed in an independent data control sample~\cite{Aaltonen:2010jr}. Contributions of {\it V}+jets and $VV$ events containing at least one incorrectly $b$-tagged light-flavored jet are determined by applying to simulated events a per-event probability, obtained from a generic event sample containing mostly light-flavored jets~\cite{Acosta:2004hw,Abulencia:2006kv}.  The efficiency of the trigger-level selection is measured in data and applied to all simulated samples. 

Because QCD multijet events with large missing transverse energy are difficult to simulate properly, a suitable model is derived solely from data; we use an independent data sample composed of events with $\Delta\varphi(\vecmet,\vec{E}_\mathrm{T}^{j_2}) <0.4$ and $50<\met<70~\gev$, consisting almost entirely of QCD multijet contributions. 
First, a $b$-tagging probability  $f_i$ is calculated separately in each \btagging subsample $i$ ($ i = $ \btagonet, \btagtl, \btagtt) by taking the ratio between tagged and pretagged events as a function of several jet- and event-related variables\,\cite{higgsmetbb}.  Then, QCD multijet kinematic distributions are determined separately for each region $i$ by weighting the untagged data in the preselection sample according to the probability  $f_i$.

The signal is modeled using \pythia for \PWprime boson mass \mwp in the range $300 \leq \mwp \leq 900$ \gevcc ~in 100 \gevcc ~increments, where the  \PWprime boson is assumed to have purely right-handed decays. As the \PWprime boson helicity does not affect analysis observables, this model is valid for both a right-handed and a left-handed \PWprime boson under the assumption of no interference with SM \PW boson production. Two scenarios are considered, depending on whether the leptonic decay mode \wplnu is  allowed or forbidden. The latter, for instance, is the case if the hypothetical right-handed neutrino \PnuR  is more massive than the \PWprime boson. The only effect of the forbidden leptonic decay mode is an increased branching fraction \wptbbr.

% \nnqcd is a 13-variable feed-forward multilayer perceptron, bearing activity-derived (\met, \mpt), angular ($\Delta\varphi(\vecmet,\vecmpt)$, angular separations between \vecmet, \vecmpt and the jet directions), and event-shape (sphericity~\cite{sphericityTensor}) observables. 
As an intermediate background-rejection step, an artificial neural network, \nnqcd, is employed to separate the dominant QCD multijet background from signal and other backgrounds. 
\nnqcd is trained using event observables (\met, \mpt \cite{Bentivegna:2012zj}), angular observables ($\Delta\varphi(\vecmet,\vecmpt)$, $\Delta \varphi (\vecmet,\vec{E}_\mathrm{T}^{j_i})$, $\Delta \varphi(\vecmpt,\vec{E}_\mathrm{T}^{j_i})$ and other topological informations such as sphericity~\cite{sphericityTensor}. 
% Single top \schannel events from simulation are used as the signal sample in the training of \nnqcd. 
As the final-state topologies for a $W'$ boson decaying to a top-bottom quark pair and $s$-channel single-top-quark production are similar, we employ the same \nnqcd function constructed to separate $W$+jets events from background in the $s$-channel single-top-quark observation\,\cite{CDF:2014uma}. No information on the $W'$ boson mass is included in the training sample in order to ensure consistent performance in QCD multijet background separation across the whole \PWprime-boson-mass range under study.

% The same \nnqcd employed in the $s$-channel single-top-quark search is used in this analysis. 
% This choice exploits the similar final state topology and kinematical distributions between $s$-channel single-top-quark production and the signal, while at the same time ensuring a consistent performance in QCD multijet background separation across the whole \PWprime mass hypothesis range, as no information on the resonant \PWprime mass is present in the training event sample, and no related bias is introduced in the training procedure.
% Figure \ref{fig:qcdnn} shows the \nnqcd output distribution for events in the \btagtt subsample. 
The events must satisfy a minimum \nnqcd requirement to maximize sensitivity to single-top-quark $s$-channel production, which is kinematically very similar to $W^{\prime}$ production at threshold. The surviving events constitute the \emph{signal} region. To determine the appropriate normalization of QCD events in each analysis subsample, we derive a scale factor in the region composed by the rejected events. Tables\,\ref{tab:yields1} and \,\ref{tab:yields2} show the event yields after the full selection.

%Machine-learning techniques are further applied to derive a final discriminant to distinguish each \PWprime-boson mass hypothesis from the remaining backgrounds. 
We use two additional neural networks, \nnnont\ and \nntt, to classify events that satisfy the minimum requirement on the \nnnonw\ output variable. The first neural network, \nnnont, is trained to separate the \PWprime boson signal from {\it V}+jets and the remaining QCD backgrounds. In the training, a simulated $W'$-boson signal is used, while the background sample consists of pretag data that satisfy the requirement on \nnnonw, reweighted by the tag-rate probability. The second neural network, \nntt, is trained to separate \PWprime boson from $t\bar{t}$ production using simulated samples. Variables that describe the energy and momentum flow in the detector and angular variables are used in the training of the \nnnont\ and \nntt\ discriminants.  The final discriminant, \nnsig, is defined as the quadrature sum of the \nnnont\ and \nntt\ output variables, both multiplied by an appropriate weight optimized to improve the expected sensitivity in each analysis subsample.
Figure~\ref{fig:nnsig1} shows the expected and observed shapes of the \nnsig\ output variable for several subsamples, with the shape corresponding to the 300\,GeV/$c^2$ \PWprime hypothesis overlaid.
% The transverse invariant mass of the \met and all jets, \mmetj, is used to discriminate the \wptb signal from the remaining backgrounds. This variable corresponds to the reconstructed transverse invariant mass of the \PWprime in events where the jets originate from \Pqb quarks, and both the lepton and the neutrino originated from the \PW boson are not detected and contribute to the \met.
% Figure \ref{fig:mvj123} shows the \mmetj distribution for events in the \btagtt subsample. 
%
% To validate the background modeling, we compare tagged data and the corresponding combined background prediction 
% % in several control regions~\cite{crs} 
% for various kinematic, angular, and event-shape variables. 

% ======================================================================
% \section{Background Rejection}
% ======================================================================
%
%%% yields table with the ruledtabular environment
\begin{table}[htbp]
  \begin{center}
    \newcolumntype{d}{D{.}{.}{1}}
%     \caption{\label{sig2}Number of predicted and observed events in the two-jet signal region in the subsample with exactly one tight \textsc{\hobit} tagged jet (1T), one tight and one loose \textsc{\hobit} tagged jet (TL) and two tight \textsc{\hobit} tagged jets (TT). The uncertainties in the predicted numbers of events are due to the theoretical-cross-section uncertainties and to the uncertainties on signal and background modeling. Both the uncertainties and the central values are those given by the fit to the data with theory constraints.}
\caption{\label{tab:yields1}Numbers of expected and observed two-jet events with and without identified leptons, combined, in the 1T, TL, and TT subsamples. The uncertainties on the expected numbers of events are due to the theoretical and experimental uncertainties on signal and background modeling. Expected numbers of events for a right-handed \PWprime boson with SM-like couplings and a mass of 300\,GeV/$c^2$ are shown.}
  \begin{tabular}{l*{3}{c}}
    	\hline
      \hline
      Category                 & 1T & TL & TT \\ 
      \hline           
$s$-ch. single top         & \, $98     \pm  10            $ & $ 36.4  \pm  3.8     $ & $ 46.1  \pm  4.3   $ \\
$t$-ch. single top          &  $167     \pm  24            $ & $ ~7.3  \pm  1.1       $ & $~ 7.9  \pm  1.1  $ \\
$t \bar t$                    &  $457  \pm  32                $ & $ 140.9  \pm  11.1$ & $ 177.4  \pm  11.7 $ \\
$VV$                          &  $259  \pm  18               $ & $ 28.5  \pm  2.0    $ & $ 27.0  \pm  2.0 $ \\ 
$VH$                          &  $14   \pm  1                   $ & $ ~5.4  \pm  0.5    $ & $ ~7.2  \pm  0.5 $ \\
$V$+jets                     &  $3473  \pm  901          $ & $ 236.4  \pm  61.1$ & $ 156.7  \pm  38.7 $ \\
QCD                            &  $2766  \pm  103          $ & $ 220.0  \pm  16.8   $ & $ 101.5  \pm  12.2 $ \\ 
\hline
Total background     &  $7235  \pm  908          $ & $ 674.3  \pm  64.2   $ & $ 524.5  \pm  43.0 $ \\
\PWprime (300\,GeV/$c^2$) &  $156  \pm 10 $ & $ 59.9  \pm  4.6     $ & $ 84.6  \pm  7.9  $ \\
Observed                    &  7128                                &    680                       &  507     \\ 
       \hline
      \hline
    \end{tabular}        
  \end{center}
\end{table}
\begin{table}[htbp]
  \begin{center}
    \newcolumntype{d}{D{.}{.}{1}}
    \caption{\label{tab:yields2} Same as in Table I but for three-jet events.}
%      
 %    Numbers of predicted and observed three-jet events with and without identified leptons, in the 1T, TL, and TT subsamples.  The uncertainties on the predicted numbers of events are due to the theoretical and experimental uncertainties on signal and background modeling. Expected numbers of events for a right-handed \PWprime boson with SM-like couplings and 300\,GeV/$c^2$ mass are shown.}
  \begin{tabular}{l*{3}{c}}
    	\hline
      \hline
      Category                 & 1T & TL & TT \\ 
      \hline           
$s$-ch. single top             &  50   $ \pm $ 5    &    13.3 $ \pm $ 1.5 ~          &   16.2  $ \pm $ 1.6  ~ \\
$t$-ch. single top              &  ~ 91 $ \pm $ 14    &   5.8 $ \pm $   0.9              &   6.9 $ \pm $ 1.0  \\ 
$t \bar t$                        &900  $ \pm $ 65      &  148.2 $\pm$   11.6 ~      & 161.6 $\pm$  10.5 ~ \\ 
$VV$                             &106$ \pm $ 8~        &  9.7 $ \pm $ 0.7                & 7.8 $ \pm $ 0.6 \\ 
$VH$                             &  ~6  $ \pm $ 1          &  1.7 $ \pm $ 0.2               &  2.1 $ \pm $ 0.2 \\ 
$V$+jets                       &  1360 $\pm$ 357    &  80.6 $ \pm $ 21.2          &  51.6 $ \pm $ 13.4 \\
QCD                              & 1261 $\pm$  64~   & 92.8 $\pm$  9.4        ~     & 31.8 $\pm$  4.6 ~ \\ 
\hline 
Total background & 3774 $ \pm $ 369         &  352  $\pm$  26.3    & 278 $ \pm $ 17.5  \\
\PWprime (300\,GeV/$c^2$)  &  80 $ \pm $ 5  &  23.5 $ \pm $ 1.9   ~        &   28.8 $ \pm $ 3.0 ~ \\ 
Observed                     &  3613                         &  388                              &  274    \\
      \hline
      \hline
    \end{tabular}        
  \end{center}
\end{table}
%
%
% \begin{table}
% 	\begin{ruledtabular}
% 		\begin{tabular}{l D{r}{\pm}{3} D{r}{\pm}{3} D{r}{\pm}{3}}
% 			& \multicolumn{1}{c}{\btagonet} & \multicolumn{1}{c}{\btagtl} & \multicolumn{1}{c}{\btagtt}  \\
% 			\hline
% 			\multijet         & 1871r113 & 238r26 & 71r10 \\ 
% 			\wzjets           & 3757r647 & 137r25 & 94r17 \\ 
% 			Diboson           & 239 r29  & 25r3  & 24r3  \\ 
% 			\ttbar            & 690 r73  & 142r17 & 151r18 \\ 
% 			Single top        & 272 r48  & 43r8  & 51r9  \\ 
% 			\hline
% 			Total Background  & 6830r663 & 584r41 & 392r28 \\ 
% 			\hline
% 			Observed          & \multicolumn{1}{c}{6815}     & \multicolumn{1}{c}{620}    & \multicolumn{1}{c}{405} \\
% 			\hline
% 			% \multirow{2}{*}{\mwp = 300 \gev} & 96r30  & 34r11 & 45r14 \\
% 			\mwp = 300 \gev & 96r30  & 34r11 & 45r14 \\
% 			\vspace{-0.2em}(\xsbr = 1 pb) & & &  \\
% 		\end{tabular}
% 	\end{ruledtabular}
% 	\caption{Expected and observed event yields in signal region defined by the \nnqcd $>$ 0.45 requirement in the three \btagging regions \btagonet, \btagtl, \btagtt. The uncertainties in the expected number of events are due to the uncertainties on the theoretical cross section and to the uncertainties on signal and background modeling. 
% 	% Templates for background processes are normalized to their expected values. 
% 	Expected number of events for one choice of \PWprime mass and \wptbxsbr is also shown.}
% 	\label{tab:yields}
% \end{table}
%
%
\begin{figure*}[!htb]
\begin{center}
%  \begin{minipage}[b]{0.35\linewidth}
  %  \centering
    \includegraphics[width=0.325\linewidth]{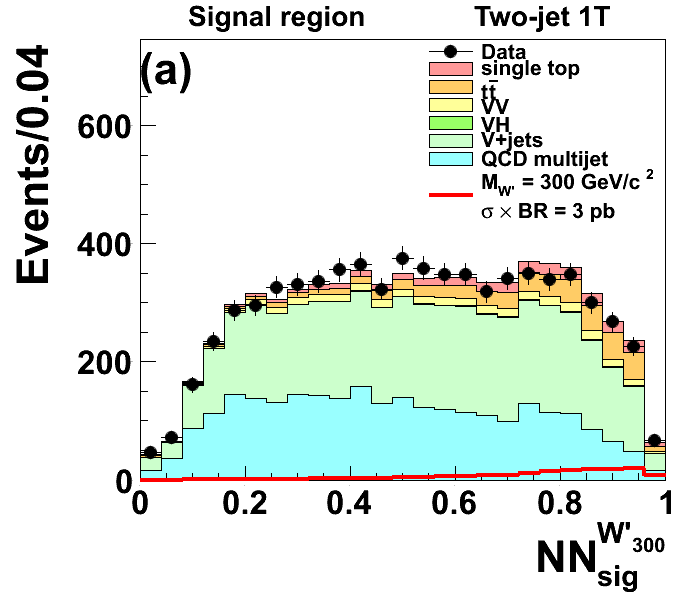}
 % \end{minipage}
 % \hspace{0.1cm}
  %\begin{minipage}[b]{0.35\linewidth}
   % \centering
    \includegraphics[width=0.325\linewidth]{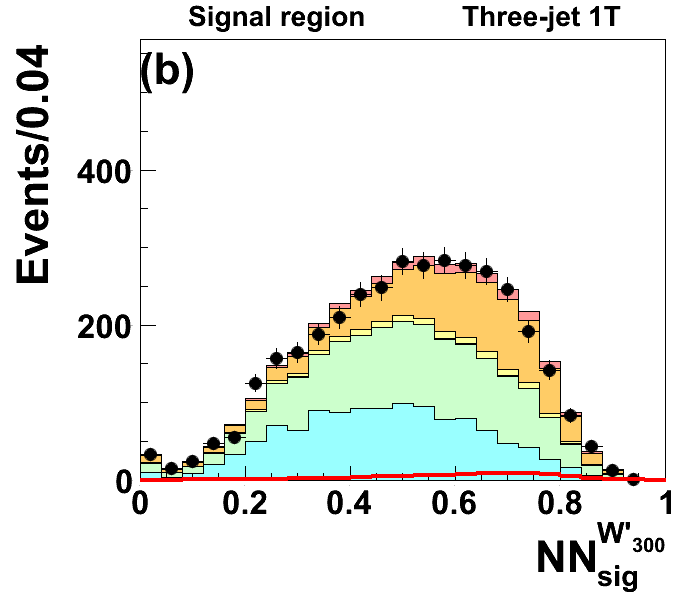}
 % \end{minipage}\\
% \hspace{0.1cm}
%  \begin{minipage}[b]{0.35\linewidth}
   %  \centering
    \includegraphics[width=0.325\linewidth]{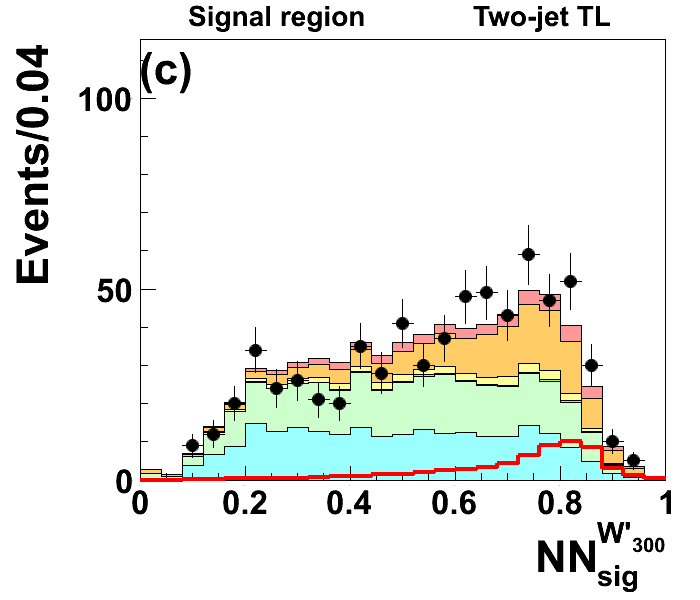}
%  \end{minipage}
%\hspace{0.1cm}
%  \begin{minipage}[b]{0.35\linewidth}
 %   \centering
    \includegraphics[width=0.325\linewidth]{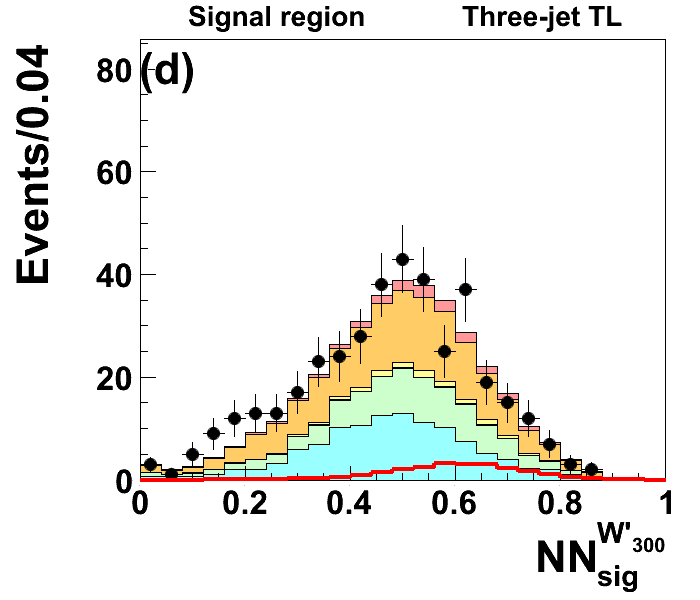}
 % \end{minipage}\\
%   \hspace{0.1cm}
 % \begin{minipage}[b]{0.35\linewidth}
   % \centering
    \includegraphics[width=0.325\linewidth]{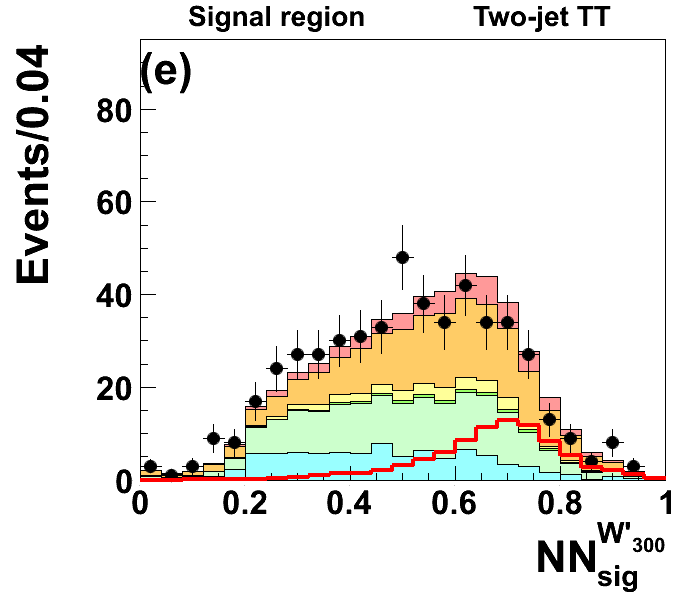}
 % \end{minipage}
%\hspace{0.1cm}
 % \begin{minipage}[b]{0.35\linewidth}
   % \centering
    \includegraphics[width=0.325\linewidth]{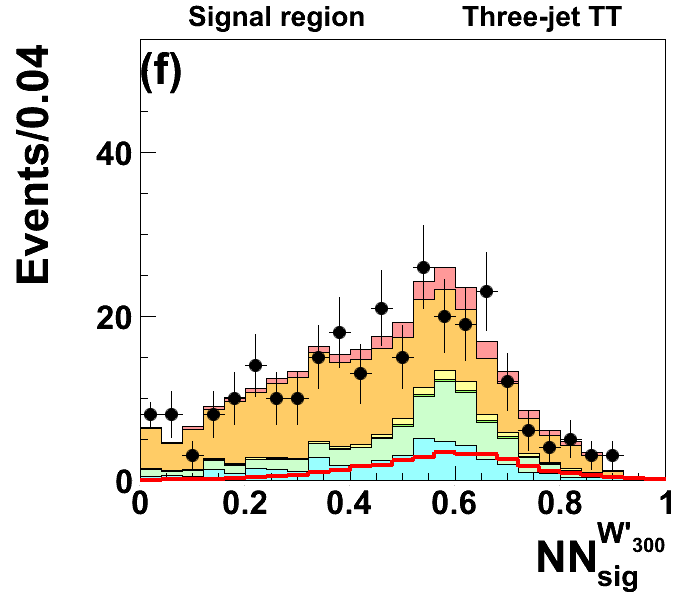}
    \end{center}
 % \end{minipage}
\caption{Expected and observed final discriminant distributions in the signal region. The distribution for a $W'$ boson with 300 GeV/\,$c^2$ mass and SM couplings is overlaid. The signal is normalized to a cross section times branching ratio of 3\,pb. From left to right, from top to bottom: 1T two-jet (a), 1T three-jet (b),  TL two-jet (c), TL three-jet (d), TT two-jet (e) and TT (f) three-jet event subsamples.}
\label{fig:nnsig1}
\end{figure*}
A binned likelihood fit is performed to probe a \wptb signal in the presence of SM backgrounds. The likelihood is the product of Poisson probabilities over the bins of the \nnsig\ ~distribution. The mean number of expected events in each bin includes contributions from each background source and from the \wptb process assuming a given value of \mwp.
We employ a Bayesian likelihood~\cite{limits} with a uniform, non-negative prior probability for the \PWprime boson production cross section times branching fraction, \wptbxsbr, and truncated Gaussian priors for the uncertainties on the acceptance and shapes of the backgrounds. 
We combine the twelve signal regions of events characterized by different \Pqb-tagging content, jet multiplicity, and presence of well-identified leptons by multiplying the corresponding likelihoods and simultaneously taking into account the correlated uncertainties. 
% All systematic uncertainties except those associated with the QCD multijet are treated as fully correlated across the tagging categories. 

Systematic uncertainties include both uncertainties on template normalization and uncertainties on the shape of the \nnsig ~distribution. 
Uncertainties due to the same source are considered 100\% correlated. These uncertainties apply to both signal and backgrounds, and include luminosity measurement (6\%), \btagging efficiency (8 to 16\%), trigger efficiency (1 to 3\%), lepton identification efficiency (2\%), parton distribution functions (3\%), initial-state and final-state simulation radiation uncertainties (2\%) and up to 6\% for the jet-energy scale \cite{cdfjets}.
The uncertainties due to finite simulation sample size, and the uncertainties on the normalization of the production of \ttbar ~ (3.5\%), $t$-channel single-top quarks (6.2\%), $s$-channel single-top quarks (5\%), dibosons (6\%) from the theoretical cross-section calculations~\cite{Kidonakis:2010tc,Campbell:1999ah}, $W+c$ (23\%) from the measured cross section~\cite{Baglio:2010um,Aaltonen:2012wn}, and QCD multijet (3 to 100\%, calculated from scale factors) are not correlated.
The production rates of events with a \PW or a \PZ boson plus heavy-flavor jets are associated with a 30\% uncertainty.
% Additional uncertainties are assigned to templates for \PW plus heavy-flavor jets (\Pqb or \Pqc) to take into account uncertainties in the determination of scale factors for the 
The shapes obtained by varying the b-tagging probability $f_i$ by one standard deviation from their central values are applied as uncertainties on the shapes of the QCD background. Changes in the shape of the \nnsig\ distribution originating from jet-energy-scale uncertainties are also incorporated for processes modeled with simulations.
% \begin{figure}[!h]
% 	\centering
% 	\includegraphics[width=\columnwidth]{multijet_TL_2j_TRF_ratio.png}
%  	\caption{Alternate QCD multijet shapes obtained by varying the probability density functions $f_i$ by one standard deviation from their central values, normalized to the nominal, for the two-jet no-lepton 1T subsample.}
% 	\label{fig:label}
% \end{figure}
% \includegraphics[width=\textwidth]}
An uncertainty on the \btagging efficiency due to different performance observed in data and simulations as a function of the jet $E_T$ is applied to signal distributions.

% The uncertainties from the simulations statistics and those on the normalizations of top-quark (10\%), diboson (6\%),  {\it V}+jets (30\%), QCD QCD (1 to 3\%), and EWK mistags (20 to 65\%) production are not correlated.
% The shapes obtained by varying the $M_{\mathit{TR}}$ (mistag) probabilities by one standard deviation from their central values are applied as shape uncertainties for the QCD MJ (EWK mistags).
% The correlated uncertainties, which apply to both the signal and the EWK backgrounds, include luminosity measurement (6\%), $b$-tagging efficiency (5 to 10\%), trigger efficiency (3-5\%), lepton veto efficiency (2\%), parton distribution function (3\%), and up to 11\% for the jet-energy scale~\cite{cdfjets}. 

% We also determine the shape uncertainties on \mmetj due to the jet-energy scale and the trigger efficiency. The latter two also affect the QCD MJ background through the background subtraction procedure described above. Initial- and final-state radiation uncertainties (2 to 3\%) are applied only to the \wptb signal. 

\begin{figure}[!h]
	\centering
	\includegraphics[width=\columnwidth]{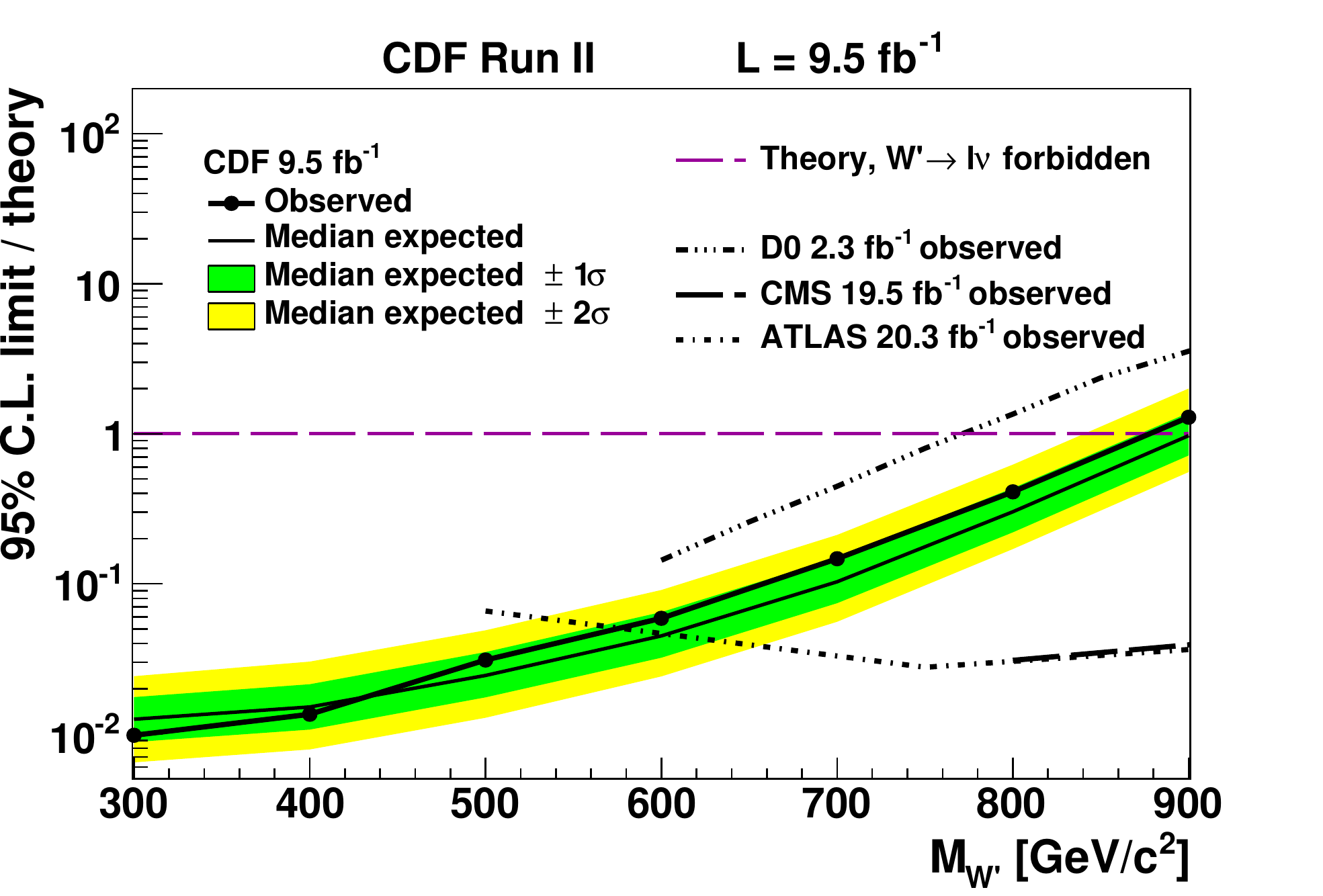}
 	\caption{Observed and expected limits on \wptbxsbr, with $\pm 1\sigma$ and $\pm 2 \sigma $ credibility intervals and theoretical predictions for a right-handed \PWprime boson with SM-like couplings in cases where the leptonic decay mode \wplnu is forbidden (dashed line). The CDF limits are compared with limits from the latest $W'$ searches from ATLAS, CMS and D0\,\cite{Aad:2014xea, Chatrchyan:2014koa, Abazov:2011xs}.}
	%Expected 95\% C.L. upper limits on the coupling strength of a right-handed \PWprime boson, normalized to the theoretical cross section times branching ratio as a function of \mwp in cases where the leptonic decay mode \wplnu is forbidden. The CDF limits are compared with limits from the latest $W'$ searches from ATLAS, CMS and D0.}
	\label{fig:limit_pb}
\end{figure}

%Observing no significant excess in the data, we
The procedure is performed for all signal mass hypotheses, obtaining 95\% Bayesian credibility (C.L.) upper limit on \wptbxsbr as functions of \mwp, using the methodology described in Ref.~\cite{Aaltonen:2010jr}. 
%in Table~\ref{tab:limits} and 
The expected and observed upper limits are shown in Fig.~\ref{fig:limit_pb}. 
The observed limits are compatible with the expectations calculated assuming that no \wptb signal is present in the data.
By comparing the limits on \wptbxsbr with the theoretical next-to-leading order calculations for a right-handed \PWprime boson with SM-like couplings \cite{PhysRevD.66.075011}, we exclude \PWprime bosons for masses less than 860 (880) \gevcc ~in cases where \wptb  decay to leptons are allowed (forbidden).

\begin{figure}[!h]
	\centering
	\includegraphics[width=\columnwidth]{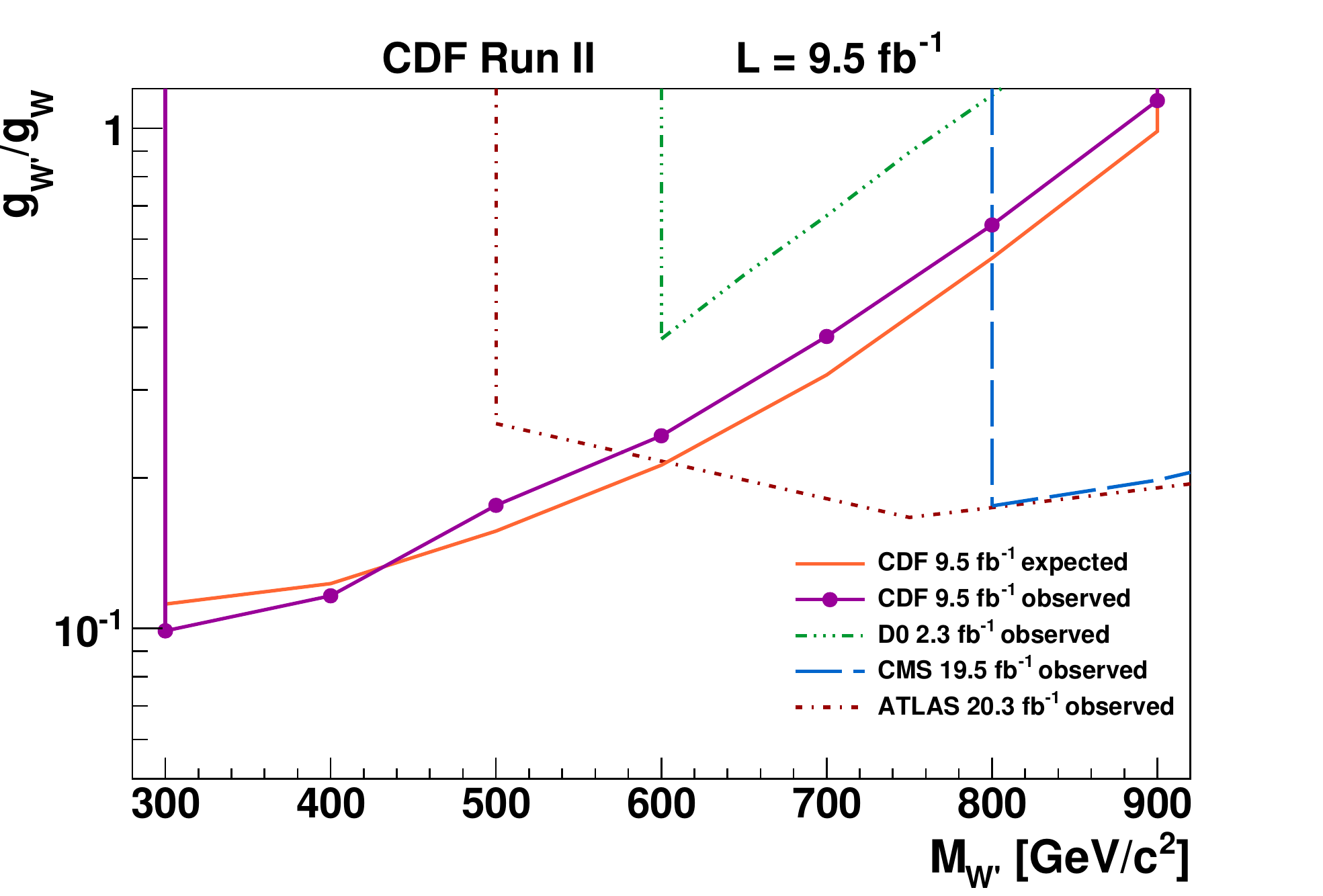}
	\caption{Observed and expected 95\% C.L. upper limits on the coupling strength of a right-handed \PWprime boson compared to the SM \PW-boson coupling, \gwp/\gsm, as functions of \mwp in cases where the leptonic decay mode \wplnu is forbidden. The region above each line is excluded. The CDF limits are compared with limits from the latest $W'$ searches from ATLAS, CMS and D0\,\cite{Aad:2014xea, Chatrchyan:2014koa, Abazov:2011xs}. The vertical part in each boundary region of the plot represents the minimum masses for which bounds are quoted}
	\label{fig:limit_wr}
\end{figure}

For a simple \schannel-production model with effective coupling \gwp, and assuming that couplings to light and heavy quarks are identical, the cross section is proportional to $\gwp^2$. By relaxing the assumption of universal weak coupling, the limits on the cross section are interpreted as upper limits on \gwp as functions of \mwp. The excluded region of the \gwp--\mwp plane is shown in Fig.~\ref{fig:limit_wr}, with \gwp expressed in units of the SM weak couplings, \gsm. For a \PWprime boson with a mass of  300 \gevcc, the effective coupling is constrained at the 95\% C.L. to be less than 10\% of the \PW boson coupling.
%
%
% ======================================================================
%\section{Results}
% ======================================================================
%
% \begin{table}
% 	\renewcommand{\arraystretch}{1.5}
% 	\setlength{\tabcolsep}{6pt}
% 	\begin{tabular}{ccc}
% 		\hline\hline
% 		\mwp [\gevcsq] & Expected $^{+1\sigma}_{-1\sigma}$ [pb] & Observed [pb] \\
% 		\hline
% 		200            & $5.40^{+3.30}_{-2.47}$            & 6.69     \\
% 		300            & $1.13^{+0.60}_{-0.38}$            & 1.42     \\
% 		400            & $0.70^{+0.33}_{-0.23}$            & 0.90     \\
% 		500            & $0.49^{+0.25}_{-0.16}$            & 0.53     \\
% 		600            & $0.31^{+0.14}_{-0.09}$            & 0.32     \\
% 		700            & $0.22^{+0.10}_{-0.07}$            & 0.23     \\
% 		800            & $0.21^{+0.10}_{-0.06}$            & 0.22    \\
% 		900            & $0.21^{+0.10}_{-0.07}$            & 0.22     \\
% 		\hline\hline
% 	\end{tabular}
% 	\caption{Expected and observed 95\% C.L. limits on \wptbxsbr as a function of \PWprime mass \mwp.}
% 	\label{tab:limits}
% \end{table}	

%\begin{figure}[!h]
%	\centering
%	\includegraphics[width=\columnwidth]{plots/results-pb.pdf}
%	\caption{Observed and expected limits on \wptbxsbr, with $\pm 1\sigma$ and $\pm 2 \sigma $ confidence intervals and theoretical predictions for a right-handed \PWprime boson with SM-like couplings in cases where the leptonic decay mode \wplnu is allowed (solid line) or forbidden (dashed).}
%	\label{fig:limit_pb}
%\end{figure}
%
In conclusion, we perform a search for a massive resonance decaying to $t b$  with the full \cdftwo data set, corresponding to an integrated luminosity of 9.5 \invfb. The data are  consistent with the background-only hypothesis, and upper limits are set on the production cross-section times branching ratio at the 95\% Bayesian credibility. For a specific benchmark model (left-right-symmetric SM extension), in cases where the \wptb-leptonic-decay mode is allowed (forbidden), we exclude \PWprime bosons with masses lower than 860 (880) \gevcc. For masses smaller than approximately 600 \gevcc, this search yields the most constraining limits to date on narrow $t b$-resonance production. 
\begin{acknowledgments}
% We thank T.\ Tait, S.\ Jung, W.\ Bernreuther, and Z.-G.\ Si for their
% assistance in preparing the theoretical models and calculations used in this
% Letter, and T.\ Rizzo for helpful conversations.  We also thank the development
% teams of \textsc{scipy}, \textsc{py}, \textsc{matplotlib}, and
% \textsc{ipython} for their useful
% tools~\cite{scipy,*pytables,*matplotlib,*ipython}.

We thank the Fermilab staff and the technical staffs of the
participating institutions for their vital contributions. This work
was supported by the U.S. Department of Energy and National Science
Foundation; the Italian Istituto Nazionale di Fisica Nucleare; the
Ministry of Education, Culture, Sports, Science and Technology of
Japan; the Natural Sciences and Engineering Research Council of
Canada; the National Science Council of the Republic of China; the
Swiss National Science Foundation; the A.P. Sloan Foundation; the
Bundesministerium f\"ur Bildung und Forschung, Germany; the Korean
World Class University Program, the National Research Foundation of
Korea; the Science and Technology Facilities Council and the Royal
Society, United Kingdom; the Russian Foundation for Basic Research;
the Ministerio de Ciencia e Innovaci\'{o}n, and Programa
Consolider-Ingenio 2010, Spain; the Slovak R\&D Agency; the Academy
of Finland; the Australian Research Council (ARC); and the EU community
Marie Curie Fellowship Contract No. 302103.

\end{acknowledgments}
\bibliographystyle{apsrev4-1-JHEPfix}
\bibliography{biblio}

\begin{thebibliography}{37}
\bibitem [{cdf()}]{cdfinfos}%
  \BibitemOpen
  \href@noop {} {}\bibinfo {note} {\cdftwo uses a cylindrical coordinate system
  with azimuthal angle $\varphi$, polar angle $\theta$ measured with respect to
  the positive $z$ direction along the proton beam, and distance $r$ measured
  from the beamline. The pseudorapidity, transverse energy, and transverse
  momentum are defined as $\eta=-\ln\left[\tan(\frac{\theta}{2})\right]$,
  $E_{T}=E\sin{\theta}$, and $p_{T}=p\sin{\theta}$, respectively, where $E$ and
  $p$ are the energy and momentum of an outgoing particle. The missing
  transverse energy \vecmet is defined by $\vecmet =-\sum_iE_T^i{\hat n}_i$,
  where ${\hat n}_i$ is a unit vector perpendicular to the beam axis that
  points to the $i$th calorimeter tower ($\met=|\vecmet|$). The missing
  transverse momentum \mpt ~is analogously defined starting from the momenta of
  reconstructed charged particles as measured by the spectrometer.}\BibitemShut
  {Stop}%
\bibitem [{sph()}]{sphericityTensor}%
  \BibitemOpen
  \href@noop {} {}\bibinfo {note} {The event sphericity is defined by
  $S=\frac{3}{2}(\lambda_2+\lambda_3)$, where the sphericity tensor is
  \mbox{$S^{\alpha\beta} = (\sum_i p_i^\alpha p_i^\beta) / (\sum_i p_i^2)$} and
  $\lambda_1>\lambda_2>\lambda_3$ are its three eigenvalues and satisfy
  $\lambda_1+\lambda_2+\lambda_3=1$. The index $i$ refers to each jet in the
  event.}\BibitemShut {Stop}%
\end{thebibliography}

\end{document}